\documentstyle[12pt,epsf]{article}
\begin{document}
\def \beq{\begin{equation}}
\def \eeq{\end{equation}}
\def \g{{\rm GeV}}
\def \msbar{\overline{\rm MS}}
\def \s{\hat s^2}
\def \seff{\sin^2 \theta_{\rm eff}^{\rm lept}}
\rightline{EFI-97-18}
\rightline{hep-ph/9704331}
\rightline{April 1997}
\vspace{0.3in}
\centerline{\bf NEW DEVELOPMENTS IN PRECISION ELECTROWEAK PHYSICS
\footnote{To be submitted to Comments on Nuclear and Particle Physics}}
\vspace{0.3in}
\centerline{\it Jonathan L. Rosner}
\centerline{\it Enrico Fermi Institute and Department of Physics}
\centerline{\it University of Chicago, Chicago, IL 60637}
\bigskip

\centerline{\bf ABSTRACT}
\medskip
\begin{quote}
Our picture of the electroweak interactions continues to improve, with ever
more precise constraints on the masses of the Higgs boson(s) and on
non-standard physics.  Some recent developments include: (a) a calculation of
higher-order effects on the weak mixing angle which reduces scheme-dependence,
(b) a long-awaited improvement in the experimental accuracy of measurement of
parity violation in atomic cesium, and (c) improvements in top quark and $W$
mass measurements from Fermilab and $W$ mass measurements from LEP. 
\end{quote}
\bigskip

\centerline{PACS Categories: 11.30.Er, 12.15.Ji, 12.15.Mm, 12.60.Cn}
\bigskip

\centerline{\bf INTRODUCTION}
\bigskip

Numerous measurements not only support the unified theory of weak and
electromagnetic interactions \cite{GWS}, but provide constraints on the
symmetry-breaking (``Higgs") sector of the theory and on possible extensions of
standard-model physics.  A streamlined language for these constraints was
provided several years ago by Peskin and Takeuchi \cite{PT}.  Within the past
year, enough has taken place that it is worth updating some recent analyses
\cite{APV,Cargese}. For several reasons, the constraints on new heavy fermions
in the electroweak sector and upper limits on the Higgs boson mass have become
somewhat stronger. In the absence of new heavy fermions, one can finally begin
to place a significant upper limit on the Higgs boson mass. However, our limit
will be somewhat more conservative than some which are quoted in recent
literature. 

The developments since the summer conferences of 1996 include: (a) a
calculation by Degrassi, Gambino, and Sirlin \cite{DGS} of higher-order effects
on the weak mixing angle which reduces scheme-dependence, (b) a long-awaited
improvement in the experimental accuracy of measurement of atomic parity
violation (APV) in cesium \cite{NewCs}, and (c) improvements in top quark and
$W$ mass measurements from Fermilab \cite{Dzerot,newCDFW,PRLD0W,newD0W} and LEP
\cite{newLEPW}. A consistent picture of electroweak parameters now includes
determination of the weak mixing angle $\sin^2 \theta$ to about 0.3\% and of
the top quark mass to about 3\%.  Other parameters (such as the $W$ mass and
the weak charge of the cesium nucleus) are specified in terms of these
quantities, and, remarkably, agree with the predictions at present levels of
accuracy.  The present article discusses what level of agreement is
significant, and what level of disagreement would point the way to new physics.

In Section II we review notation and conventions.  Section III is devoted to
the effects of the calculation of Ref.~\cite{DGS}.  The experimental inputs
are described in Section IV, while Section V gives the results of a
simultaneous fit to the observables.  Possible future improvements in
measurements and their theoretical interpretation are treated in Section VI,
while Section VII summarizes.
\bigskip

\centerline{\bf II.  NOTATION AND CONVENTIONS}
\bigskip

The strengths of electroweak interactions are specified by two coupling
constants, $g$ and $g'$, corresponding to the SU(2) and U(1) gauge groups, and
a Higgs boson vacuum expectation value $v = 2^{-1/4} G_F^{-1/2} \simeq 246$
GeV, where $G_F = 1.16639(2) \times 10^{-5}$ is the Fermi coupling constant.
Separate measurements of $g$ and $g'$ are not available, since -- in contrast
to the unit of electric charge $e$ -- neither is probed by a long-range
interaction.  However, the two couplings are related to $e$ through $e =
gg'/(g^2 + g'^2)^{1/2}$.  It is most convenient to express all couplings in
terms of their values at some convenient mass scale. Since so many measurements
have taken place at the $Z$ peak, this scale is taken to be $M_Z$.
Extrapolation of the fine-structure constant $\alpha = e^2/4 \pi = 1/137.036$
to $M_Z$ then leads to the estimate \cite{alpha} $\alpha^{-1} (M_Z) = 128.9 \pm
0.1$. 

The mass of the $Z$ boson itself involves another combination of the parameters
$g$, $g'$, and $G_F$.  To lowest order, $G_F/\sqrt{2} = (g^2 + g'^2)/8 m_Z^2$.
The measurement of $M_Z$ at LEP has become so precise that one must correct for
distortions of the LEP ring by earth tides and changes in ground water, and
effects induced by passage of the TGV train between Geneva and Paris.  The
result \cite{EWWG} is $M_Z = 91.1863 \pm 0.0020 ~\g /c^2$. 

Thus, knowing $G_F$, $\alpha(M_Z)$, and $M_Z$, we should be in a position to
predict all electroweak observables, including the weak mixing angle $\theta$,
where $\sin^2 \theta = e^2/g^2$.  For example, one expects $M_W = M_Z \cos
\theta$.  This lowest-order relation, and most others, are affected by fermion
loops in gauge boson propagators.  The most important such effect \cite{Tini}
is due to the heavy top quark, modifying the relation between coupling contants
and $M_Z$ to read
\beq
\frac{G_F}{\sqrt{2}} \rho = \frac{g^2 + g'^2}{8 m_Z^2}~~,~~~
\rho \simeq 1 + \frac{3 G_F m_t^2}{8 \pi^2 \sqrt{2}}~~~.
\eeq
Here we have neglected all quark masses except $m_t$.  The Collider Detector
Facility (CDF) Collaboration at Fermilab \cite{Watop} reported $m_t = 176.8
\pm 6.5~\g /c^2$ at the summer conferences in 1996, while a new value from
the D0 Collaboration \cite{Dzerot} has recently appeared: $m_t = 173.3 \pm 5.6
\pm 6.2~\g/c^2$.  The average of the two measurements is \cite{topav} $m_t =
175.5 \pm 5.5~\g/c^2$, leading to about a 1\% upward correction to $\rho$.
Many electroweak measurements have far surpassed this accuracy and thus are
sensitive to the top quark mass.

The Higgs boson mass $M_H$ also affects electroweak observables through loops,
but only logarithmically.  Thus one has to know both $m_t$ and $M_H$ in order
to use the values of $G_F$, $\alpha(M_Z)$, and $M_Z$ to predict other
quantities.  The effects of $m_t$, $M_H$, and a number of other sources of new
physics (``oblique corrections'') whose effects are felt mainly through loops
in gauge boson propagators can be gathered into three parameters called
$S$, $T$, and $U$ by Peskin and Takeuchi \cite{PT}.  Each observable can be
expressed as a ``nominal'' value for fixed $G_F$, $\alpha(M_Z)$, $M_Z$,
$m_t$, and $M_H$, plus a linear combination of $S$, $T$, and $U$.  The only
place $U$ appears is in the expression for the $W$ mass.  A pair of parameters
equivalent to $S$ and $U$ \cite{MR} is $S_Z \equiv S$ and $S_W \equiv S + U$.
We shall take as nominal values $m_t = 175~\g/c^2$ and $M_H = 300~\g/c^2$.

When $m_t$ and $M_H$ deviate from their nominal values, and when new heavy
fermions $U$ and $D$ with $N_C$ colors and charges $Q_U$ and $Q_D$ are present,
one finds the following contributions \cite{KL}: 

$$
T \simeq \frac{3}{16 \pi \sin^2 \theta M_W^2} \left[ m_t^2 -
(175 ~{\rm GeV})^2 + \sum \frac{N_C}{3} \left(m_U^2 + m_D^2 - \frac{2 m_U^2
m_D^2}{m_U^2 - m_D^2} \ln \frac{m_U^2}{m_D^2} \right)  \right]
$$
\beq \label{eqn:t}
- \frac{3}{8 \pi \cos^2 \theta} \ln \frac{M_H}{300~{\rm GeV}} ~~~, 
\eeq
where $\Delta \rho \simeq \alpha T$;
\beq \label{eqn:sz}
S_Z = \frac{1}{6 \pi} \left [ \ln \frac{M_H}{300~\g/c^2} - 2 \ln
\frac{m_t}{175~\g/c^2} + \sum N_C \left ( 1 - 4 \overline Q \ln
\frac{m_U}{m_D} \right ) \right ] ~~~,
\eeq
\beq \label{eqn:sw}
S_W = \frac{1}{6 \pi} \left [ \ln \frac{M_H}{300 ~\g/c^2} + 4 \ln
\frac{m_t}{175~\g/c^2} + \sum N_C \left ( 1 - 4 Q_D \ln \frac{m_U}{m_D}
\right ) \right ]~~.
\eeq
The expressions are written for doublets of fermions with $N_C$ colors and $m_U
\geq m_D \gg m_Z$, while $\overline Q \equiv (Q_U + Q_D ) /2$. The sums are
taken over all doublets of new fermions. The leading-logarithm expressions are
of limited validity for $M_H$ and $m_t$ far from their nominal values.  We do
not consider multi-Higgs models here; a discussion can be found in \cite{AKG}. 

Most electroweak observables can be written as homogeneous functions of $\rho$
(typically of degree 0, 1, or 2) times linear functions of $\sin^2 \theta$
(which itself is a linear function of $S$ and $T$).  Since $\Delta \rho =
\alpha T$, it is then straightforward to evaluate the coefficients of $S$ and
$T$ in such observables \cite{PT,MR}.  Examples of these coefficients will be
given in Sec.~V when we discuss fits to the data. 
\bigskip

\centerline{\bf III.  DEFINITIONS OF WEAK MIXING ANGLE}
\bigskip

Several definitions of the weak mixing angle have appeared in the literature. 
The {\it on-shell} scheme \cite{Sirlin} defines a value of $\theta$ valid in
the presence of loop corrections via the tree-level relation $M_W = M_Z \cos
\theta$.  The $\msbar$ (modified-minimal-subtraction) scheme is more directly
related to the ratio of coupling constants: $\s \equiv \sin^2 \theta_{\msbar}
\equiv (e^2/g^2)_{\msbar}$ with a particular prescription for removing
divergences in loop corrections.  Finally, $\seff$ is the effective value of
$\sin^2 \theta$ measured in leptonic asymmetries at the $Z$, which probe the
ratio of the vector and axial vector couplings of leptons to the $Z:~g_V =
-(1/4) + \seff,~g_A = 1/4$. 

In previous analyses \cite{APV,Cargese} we made use of a connection between the
nominal value of $\s$ as quoted by DeGrassi, Kniehl, and Sirlin \cite{DKS} for
the difference \cite{GS} $\seff - \s = 0.0003$, to quote $\s = 0.2315$ for $m_t
= 175~\g/c^2$, $M_H = 300~\g/c^2$.  This difference has now diminished to
0.0001 \cite{DGS}, and the nominal value of $\s$ is now quoted as 0.23200.  The
net effect of these changes is to raise the nominal value of $\seff$ from
0.2318 to 0.23211.  As we will see in Sec.~V, the experimental values of
$\seff$ measured at LEP and SLC then imply a more negative value of $S$ when
other constraints are taken into account.  This shift favors a lower Higgs
boson mass for a fixed value of $m_t$. 
\bigskip

\centerline{\bf IV.  EXPERIMENTAL INPUTS}
\bigskip

\leftline{\bf A.  Improved measurement of parity violation in atomic cesium}
\bigskip

An early prediction of the electroweak theory was the violation of parity in
atomic transitions through the mixing of opposite-parity levels induced by the
weak neutral current.  Within the past 10 years increasingly precise
experiments \cite{CW,Bi,Pb,TlS,TlO} have been performed in various systems, and
all (so far) have agreed with electroweak predictions.  At the time of a recent
review \cite{APV}, the most precise information was provided by a measurement
in cesium \cite{CW}. The result can be quoted in terms of a weak charge $Q_W$,
which measures the strength of the (coherent) vector coupling of the $Z$ to the
nucleus: $Q_W \simeq \rho(N - Z - 4 Z \s)$.  It was found that $Q_W({\rm Cs}) =
-71.04 \pm 1.58 \pm 0.88$.  The first error is experimental, while the second
is theoretical \cite{CsthNov,CsthND}. 

When the measurements of Ref.~\cite{CW} were first reported, theoretical
calculations hadn't been done well enough yet for the result to have an impact,
and the Peskin-Takeuchi $S-T$ language \cite{PT} was not yet currently in use
(though there were some intimations of it in the literature) \cite{MVS}. 

The atomic physics calculations by the Novosibirsk group in 1989 \cite{CsthNov}
and the Notre Dame group in 1990 \cite{CsthND}, and the $S-T$ description by
Peskin and Takeuchi \cite{PT} changed the situation.  It had been noticed
\cite{PS,JRRC} that the weak charge was insensitive to the electroweak
radiative corrections due to the top quark mass once the $Z$ mass was
specified.  The impact of this result in the $S-T$ language \cite{MR} was
realized at a workshop at Snowmass in June of 1990, where it was found that the
weak charge for cesium (and atoms in that range of A and Z) was almost
independent of $T$.  Thus it took a couple of years between the reporting of
the experimental results and the full realization of their significance for
particle physics. 

The prediction \cite{MR} $Q_W({\rm Cs}) = -73.20 \pm 0.13$ is insensitive to
standard-model parameters \cite{MR,PS}; discrepancies are good indications of
new physics (such as exchange of an extra $Z$ boson).  The 1988 result in
cesium could be interpreted as implying $S = -2.6 \pm 2.3$. 

The weak charge has also been measured in atomic thallium: $Q_W({\rm Tl}) =
-114.2 \pm 1.3 \pm 3.4$ \cite{TlS} and $-120.5 \pm 3.5 \pm 4.0$ \cite{TlO}
(this number is deduced \cite{APV} from the published result), to be compared
with the theoretical estimate \cite{Tlth,PSBL} $Q_W = -116.8$. Here the first
error refers to the total experimental error, while the second refers to the
error associated with the atomic physics calculations.  Averaging the two
values, we find \cite{APV} $Q_W({\rm Tl}) = -115.0 \pm 2.1 \pm 4.0 = -115.0 \pm
4.5$, implying $S = -4.5 \pm 3.8$.  As in the case of cesium, the central value
is negative, but consistent with zero.  An accurate measurement in lead
\cite{Pb} awaits comparable progress in the theoretical calculation
\cite{Pbth}. 

A new experimental value for cesium has now appeared \cite{NewCs}: $Q_W({\rm
Cs}) = -72.11 \pm 0.27$.  The theoretical error from the atomic physics
calculations is still $\pm 0.88$, leading to a total error of $\pm 0.93$. This
result implies $S = -1.3 \pm 0.3 \pm 1.1$, a value still consistent with zero,
but with a reduced error. 
\bigskip

\leftline{\bf B.  New $W$ mass measurements}
\bigskip

1.  {\it Direct production of $W^+ W^-$ pairs at LEP.}  All four LEP
experiments reported $W$ mass measurements at the 1996 Warsaw Conference
\cite{LEPW, Blondel}, with an average value of $M_W = 80.3 \pm 0.4 \pm
0.1~\g/c^2$.  These results were based on the reaction $e^+ e^- \to W^+ W^-$
just above threshold, at a center-of-mass energy of 161 GeV.   For this energy,
the $W$ mass measurement is based mainly on comparison of the experimentally
measured cross section with the theoretical prediction. 

The LEP Collider has now been operated at c.m. energies up to 172 GeV, where
the $W$ mass must be measured by reconstructing its decay.  This entails a
larger systematic error, though the higher cross section farther above
threshold makes for a lower statistical error for a given integrated
luminosity. The combined result of all four LEP experiments \cite{newLEPW} is
now $M_W = 80.38 \pm 0.14~\g/c^2$. 

2.  {\it Production in the Tevatron.}  The CDF and D0 collaborations at
Fermilab have reported new $M_W$ values based on data obtained during Run 1B,
which ended in 1996.  The new CDF value \cite{newCDFW}, based on $W \to \mu
\nu$, is $M_W = 80.430 \pm 0.100 \pm 0.040 \pm 0.115~\g/c^2$, where the first
error is statistical, the second is due to uncertainty in the momentum scale,
and the third is the remaining systematic error.  This value of $M_W = 80.430
\pm 0.155~\g/c^2$ may be combined with previous CDF $e \nu$ and $\mu \nu$ data
to yield $M_W = 80.375 \pm 0.120~\g/c^2$. Analysis of the $W \to e \nu$ decays
from Run 1B is in progress. 

The new D0 $W\ \to e \nu$ value, based on a value recently reported
\cite{newD0W} and one published last year \cite{PRLD0W}, is $80.44 \pm
0.11~\g/c^2$. When the earlier UA2 \cite{UA2W} result of $80.36 \pm
0.37~\g/c^2$ is combined with the CDF and D0 results, taking account of
common errors, the average is $M_W = 80.41 \pm 0.09~\g/c^2$. 

3.  {\it World average of direct measurements.}  The LEP II and hadron collider
measurements may be combined to give $M_W = 80.401 \pm 0.076~\g/c^2$.

4.  {\it Indirect measurement using deep inelastic neutrino scattering.} The
ratio $r_\nu \equiv \sigma(\nu_\mu N \to \nu_\mu + \ldots)/\sigma(\nu_\mu N \to
\mu^- + \ldots)$ is very sensitive to $\sin^2 \theta$ and provides a good
measure of it.  It turns out that the $\sin^2 \theta$ and $\rho$ dependences
combine in such a way that this ratio actually depends on $m_t$ and $M_H$ in
very much the same way as does $M_W$.  As a result, measurements of $R_\nu$ can
be quoted as effective measurements of $M_W$.

A new measurement by the CCFR Collaboration at Fermilab \cite{CCFR} implies
$M_W = 80.35 \pm 0.21~\g/c^2$.  This analysis was performed for $m_t =
175~\g/c^2$ and $M_H = 150~\g/c^2$; the variation of $M_W$ with $m_t$ and $M_H$
within the region of interest may be ignored. Previous determinations
\cite{CDHS,CHARM} imply slightly lower values of $M_W$ but are hard to combine
with the CCFR value without updated analyses.  Consequently, we do not include
them in our averages. 
\bigskip

\leftline{\bf C.  Progress on the top quark}
\bigskip

As of the 1996 Warsaw Conference, CDF was reporting $m_t = 176.8 \pm 6.5
~\g/c^2$, while D0 reported $m_t = 169 \pm 11~\g/c^2$.  A new D0 result
has now appeared \cite{Dzerot}: $m_t = 173.3 \pm 5.6 \pm 6.2~\g/c^2$, leading
to a world average of $m_t = 175.5 \pm 5.5~\g/c^2$.  The result is unlikely to
change much in the next few years, as both experiments have analyzed their
full data set.
\bigskip

\leftline{\bf D.  Summary of inputs}
\bigskip

We summarize in Table 1 the experimental inputs for a simultaneous fit to
electroweak observables, performed in the spirit of Refs.~\cite{PT} and
\cite{MR}.

We do not use the following quantities in the fit:  (1)  The total width of the
$Z$, $\Gamma_Z = 2.4946 \pm 0.0027~\g$, has been used here to extract the
leptonic width \cite{EWWG}, and is not an independent quantity.  To include it
in the fit one should take account of correlations, as in the treatments of
Refs.~\cite{LLM,LE}.  The independent information which it provides is a
confirmation of the currently accepted value of $\alpha_s = 0.118 \pm 0.006$
\cite{Schmelling,Alt} and/or the absence of significant additional decay
channels of the $Z$ into new particles.  (2) The partial width $\Gamma(Z \to b
\bar b)$ can receive a contribution from a virtual top quark loop and thus is
expected to differ from $\Gamma(Z \to d \bar d)$ or $\Gamma(Z \to s \bar s)$. 
Until recently the effects of the top quark loop, which should act to suppress
the partial width somewhat, had not been detected.  The ALEPH Collaboration
\cite{AZbb}, after a silence of several years, presented results at the 1996
Warsaw Conference much more in accord with this expectation, and other LEP
experiments now appear to agree with this finding.  Nonetheless, since
$\Gamma(Z \to b \bar b)$ is still in a state of flux, we do not use it in our
fit. 

\begin{table}
\begin{center}
\caption{Electroweak observables described in fit}
\medskip
\begin{tabular}{c c c} \hline
Quantity      &   Experimental   &   Theoretical \\
              &      value       &    value      \\ \hline
$Q_W$ (Cs)    & $-72.11 \pm 0.93^{~a)} $  &  $ -73.20^{~b)} - 0.80S - 0.005T$\\
$Q_W$ (Tl)    & $-115.0 \pm 4.5^{~c)} $ &  $ -116.8^{~d)} -1.17S - 0.06T$ \\
$M_W~(\g/c^2)$ & $80.41 \pm 0.10^{~e)}$  & $80.308^{~f)} -0.29S + 0.45T$ \\
$M_W~(\g/c^2)$ & $80.38 \pm 0.14^{~g)}$  & $80.308^{~f)} -0.29S + 0.45T$ \\
$M_W~(\g/c^2)$ & $80.35 \pm 0.21^{~h)}$  & $80.308^{~f)} -0.29S + 0.45T$ \\
$\Gamma_{\ell\ell}(Z)$ (MeV) & $83.91 \pm 0.11^{~i)}$ & $83.90 -0.18S
+ 0.78T$ \\
$\seff$ & $0.23200 \pm 0.00027^{~j)}$ & $0.23211^{~k)}
 + 0.0036S - 0.0026T$ \\
$\seff$ & $0.23061 \pm 0.00047^{~l)}$ & $0.23211^{~j)}
 + 0.0036S - 0.0026T$ \\ 
$m_t~(\g/c^2)$ & $175.5 \pm 5.5$ & $175 + 241S + 82T $ \\ \hline
\end{tabular}
\end{center}
\leftline{$^{a)}$ {\small Weak charge in cesium \cite{NewCs}}}
\leftline{$^{b)}$ {\small Calculation \cite{MR} incorporating 
atomic physics corrections \cite{CsthNov,CsthND}}}
\leftline{$^{c)}$ {\small Weak charge in thallium \cite{TlS,TlO} (see text)}}
\leftline{$^{d)}$ {\small Calculation \cite{PSBL} incorporating
atomic physics corrections \cite{Tlth}}}
\leftline{$^{e)}$ {\small Average of direct hadron collider measurements}}
\leftline{$^{f)}$ {\small Including perturbative QCD corrections \cite{DGS}}}
\leftline{$^{g}$ {\small LEP II value as of March, 1997}}
\leftline{$^{h}$ {\small Value from deep inelastic neutrino scattering
 \cite{CCFR}}}
\leftline{$^{i)}$ {\small LEP average as of November, 1996 \cite{EWWG}}}
\leftline{$^{j)}$ {\small From asymmetries at LEP \cite{EWWG}}}
\leftline{$^{k)}$ {\small As calculated \cite{DGS} with correction for
relation between $\seff$ and $\s$}}
\leftline{$^{l)}$ {\small From left-right asymmetry in annihilations at
SLC \cite{SLC}}}
\end{table}

We average several quantities before fitting them.  By combining $W$ mass
measurements from hadron and $e^+ e^-$ colliders and from deep inelastic
neutrino scattering we find $M_W = 80.395 \pm 0.071~\g/c^2$.  The average of
$\seff$ from LEP and SLC gives $\seff = 0.23166 \pm 0.00058$ when we multiply
the error by a scale factor \cite{PDG} of $[\chi^2/(N-1)]^{1/2} = 2.55$ to
account for the discrepancy between the two values. [We cannot think of a
plausible fundamental reason for this discrepancy.  One could suspect
systematic errors, e.g., in polarization measurements at SLC or in the
interpretation of data at LEP to extract such quantities as $A_{FB}^b$.] Adding
a theoretical error of $\Delta \seff = \pm 0.00026$ associated with the assumed
error of $\pm 0.1$ in $\alpha^{-1}(M_Z)$, we employ the value $\seff = 0.23166
\pm 0.00064$ in our fits. 
\bigskip

\centerline{\bf V.  RESULTS OF FIT TO OBSERVABLES}
\bigskip

\begin{figure}
\centerline{\epsfysize = 2.9 in \epsffile {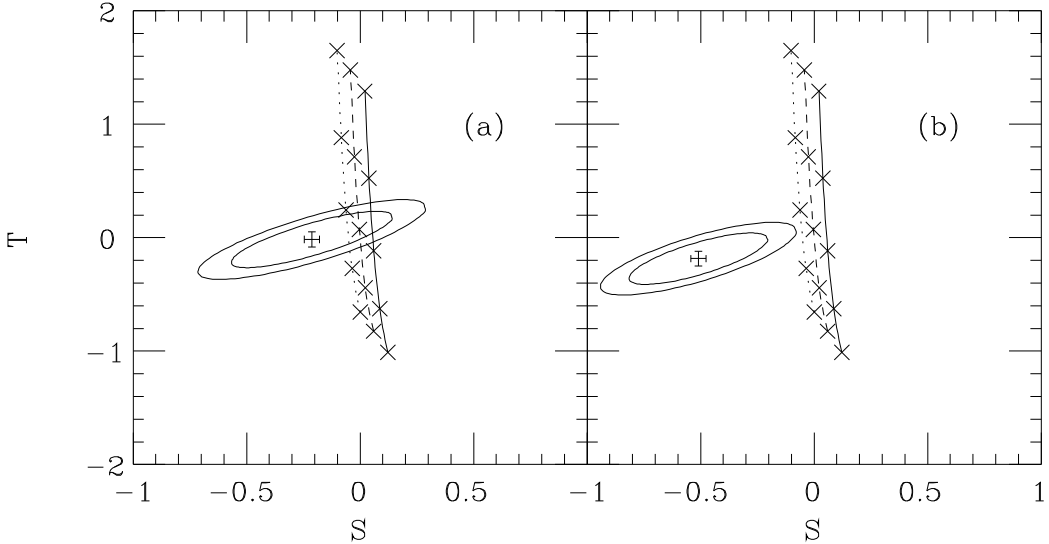}}
\caption{Allowed ranges of $S$ and $T$ at 68\% (inner ellipses) and 90\% (outer
ellipses) confidence levels, corresponding to $\chi^2 = 2.3$ and 4.6 above the
minima (crosses at center of ellipses).  Dotted, dashed, and solid lines
correspond to standard model predictions for $M_H = 100$, 300, 1000 GeV/$c^2$.
Symbols $\times$, from bottom to top, denote predictions for $m_t = 100$, 140,
180, 220, and 260 GeV/$c^2$. (a) Fit including APV experiments with present
errors; (b) errors on APV experiments reduced from $\Delta Q_W = \pm 0.93$
(present value) to $\pm 0.3$ (corresponding to a hypothetical improvement in
theoretical error), with present central values of $Q_W$ retained.} 
\end{figure}

The results can be displayed in two different ways.  In Figure 1 we show error
ellipses corresponding to the allowed ranges of the parameters $S$ and $T$
discussed in Section II for the data summarized in Table 1, but not including
the constraint due to the top quark mass.  Figure 2(a) shows the fit for this
same data set, with the additional stipulation that $m_t = 175.5 \pm
5.5~\g/c^2$. 

The present data [Fig.~1(a)] are fully consistent with electroweak predictions.
$S$ and $T$ can be viewed as free parameters pointing the way to new physics. 
Fig.~1(b), to be discussed below,  shows that improved errors on the
theoretical prediction for parity violation in atomic cesium can have a notable
effect on the error ellipses in $S$ and $T$, given present accuracy of
experiments. 

One can also view $S$ and $T$ simply as a way of parametrizing the electroweak
theory.  In this case one can include information on the top quark mass by
linearizing Eqs.~(\ref{eqn:t}) and (\ref{eqn:sz}) in $m_t - 175~\g/c^2$ and
eliminating the dependence on $\ln(M_H/300~\g/c^2)$, resulting in the relation
$m_t = 175~\g/c^2 + 241 S + 82 T$.

Some preference is shown for a particular range of Higgs boson masses when
information on the top quark mass is included, as shown in Fig.~2(a). We find
that $M_H \simeq 166 \times 2^{\pm 1}~\g/c^2$ leads to a variation of $\chi^2$
of one unit above the minimum (2.13 for 4 degrees of freedom -- the 6 data
points in Table 1 minus the two parameters $S$ and $T$). For $\Delta \chi^2 =
2.3$, the corresponding uncertainty in $M_H$ is about $(2.8)^{\pm 1}$. The
higher central value of our $M_H$ in comparison with the values
$127^{+143}_{-71}~\g/c^2$ \cite{DGS} or $124^{+125}_{-71}~\g/c^2$ \cite{LE} is
due primarily to our use of a scale factor in quoting the error on $\seff$.
Until the discrepancy between the LEP and SLC values is resolved, we regard
this as prudent.  Another point of view \cite{Blondel,LE} is to explore the
effect of ignoring one or more mildly discrepant values of $\seff$, such as the
SLC value or the value $\seff = 0.23246 \pm 0.00041$ (entering into the LEP
average) due to the forward-backward asymmetry in $e^+ e^- \to Z \to b \bar b$.
The corresponding uncertainty in the Higgs boson mass is similar to that
implied by Fig.~2(a). 

\begin{figure}
\centerline{\epsfysize = 2.9 in \epsffile {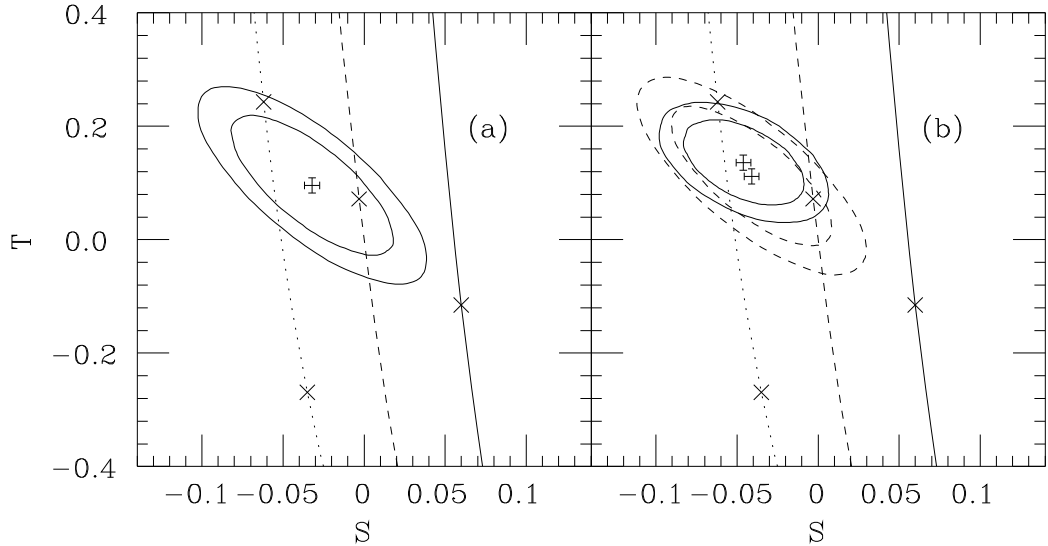}}
\caption{Magnified view of Figure 1.  Dotted, dashed, and solid lines
correspond to standard model predictions for $M_H = 100$, 300, 1000 GeV/$c^2$.
Symbols $\times$ denote predictions for $m_t = 140$ (bottom) and 180 (top)
GeV/$c^2$ on $M_H = 100$ GeV/$c^2$ curve, and for $m_t = 180$ GeV/$c^2$ on $M_H
= 300,~1000$ GeV/$c^2$ curves. (a) The constraint $m_t = 175.5 \pm 5.5~\g/c^2$
has been imposed. (b) Dashed ellipses:  same as (a) but with $\Delta Q_W({\rm
Cs}) = \pm 0.3$.  Solid ellipses: same as (a) but with $\Delta M_W = \pm 30$
MeV/$c^2$.} 
\end{figure}

In any case, our analysis now disfavors Higgs bosons heavier than $500~\g/c^2$,
unless new contributions to the parameter $T$ (beyond the top quark) exist.
That this is still a possibility can be seen in Fig.~1(a) from the solid
curve corresponding to $M_H = 1000~\g/c^2$, which passes within the 68\% c.l.
ellipse for a range of top quark masses above $200~\g/c^2$.
\bigskip 

\centerline{\bf VI.  POSSIBLE IMPROVEMENTS AND THEIR IMPACT}
\bigskip 

\leftline{\bf A.  Atomic parity violation}
\bigskip 

The present measurement can be placed in the context of present constraints on
$S$ and $T$ from high energy experiments.  As in 1988, not much can be said
until the theoretical errors shrink to match the experimental ones.  The 90\%
error ellipses in $S$ and $T$ specified by high energy experiments (primarily
those at LEP, SLC, and Fermilab) allow a range of only $\pm 0.4$ in $S$ from
the central value ($S$ close to 0).  When the total error on $S$ (experiment
plus theory) from atomic physics experiments is $\pm 0.4$, this experiment will
again provide a significant constraint in the $S-T$ context.  An example is
shown in Fig.~1(b). Here, it is assumed that the theoretical errors in
extracting $Q_W({\rm Cs})$ from experiment have been decreased so that the
total error on $Q_W$ is only $\pm 0.3$.  The central value has been assumed to
be the present one.  A significant constraint on the electroweak theory
results. 

Viewed simply as alternative ways to measure electroweak parameters, atomic
parity violation experiments have less impact than the accelerator experiments.
In Fig.~2(b) we show via the dashed curves the $S-T$ contours for an improved
determination of $Q_W({\rm Cs})$ when the top quark mass is taken into account.
The curves are indistinguishable from those in Fig.~2(a). Another illustration
of this point is provided by the discussion \cite{APV} of the impact of
measurements of isotope ratios \cite{ratio} of $Q_W$.  These turn out to be
primarily sensitive to $\seff$ and, as such, do not compete in accuracy with
LEP and SLC measurements. 

We reiterate that a crucial comparison involves Figs.~1(a) and 1(b), where the
possibility of new physics is kept open.  The key measurement is the {\it
absolute scale} of the parity-violating transition, as recently improved so
notably for cesium \cite{NewCs}.  A corresponding improvement in the atomic
theory is of high priority in order to extract a value of $Q_W$ with suitably
small errors. 
\bigskip

\leftline{\bf B.  $W$ mass}
\bigskip

The analysis in Ref.~\cite{DGS} entailed a range of Higgs boson masses from
which one could infer an allowed range $M_W = 80.367 \pm 0.048~\g/c^2$.  We
extract a range of $M_W$ in our fit by studying the variation of $\chi^2$ in a
fit to all observables except $M_W$ for a range of fixed values of $M_W$.  We
find $M_W = 80.338 \pm 0.057~\g/c^2$ for $\Delta \chi^2 = 1$ above the minimum.
Our slightly lower central value and larger uncertainty are correlated with our
slightly higher central value and larger uncertainty for $M_H$.  We may combine
our value of $M_W$ in this fit with the experimental average to obtain an
overall average $M_W = 80.360 \pm 0.044~\g/c^2$.  An earlier result
\cite{EWWG}, $M_W = 80.352 \pm 0.033$, does not incorporate all the data
considered here.  Our uncertainty is larger primarily because of the scale
factor by which we have multiplied the uncertainty in $\seff$. 

We thus conclude that a reduction of uncertainty in direct measurements of
$M_W$, e.g., to $\pm 30$ MeV/$c^2$, could match or surpass the accuracy
provided by fits to the LEP and SLD data, as long as the discrepancy persists
between the $\seff$ values measured at LEP and SLC.  To illustrate this point,
we show in Fig.~2(b) via the solid curves the effect of reducing the errors on
$M_W$ to $\pm 30$ MeV/$c^2$. 

In all of the present discussion, we have been assuming $S_W - S_Z \equiv U =
0$, which follows if there are no weak isospin-violating effects due to new
heavy particles.  If such effects are present, one may spot them first through
a discrepancy between direct measurements of $M_W$ and the predictions based on
fits to LEP and SLC data.  Present fits \cite{LE} to electroweak data suggest
$U \simeq 0.07 \pm 0.42$. 
\bigskip

\leftline{\bf C.  Top quark mass and $\seff$}
\bigskip

\begin{figure}
\centerline{\epsfysize = 2.9 in \epsffile {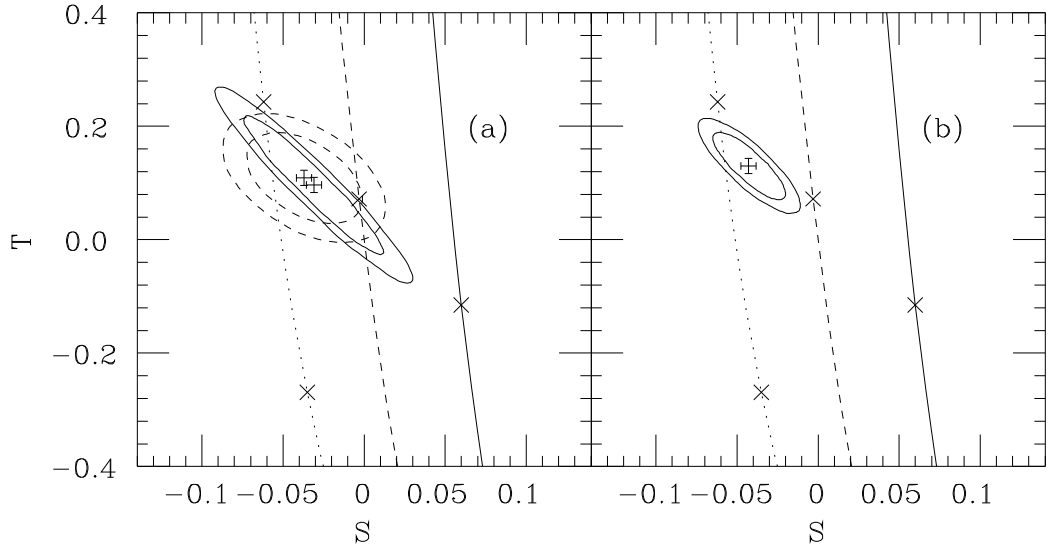}}
\caption{(a) Solid curves: same as Figure 2(a), but with $m_t = 175.5 \pm
2~\g/c^2$ assumed.  Dashed curves: same as Figure 2(a), but with $\seff =
0.23166 \pm 0.00023$ assumed. (b) Same as Figure 2(a), but with improved errors
mentioned above for $M_W$, $m_t$, and $\seff$.} 
\end{figure}

The shape of the contours in Fig.~2(a) implies that further improvement in
measurement of the top quark mass is unlikely to restrict the parameters of the
electroweak theory unless it is combined with an improved $W$ mass measurement
or a resolution of the discrepancy in measuring $\seff$.  The latter observable
is closest of all those we have considered to one which would restricts the
semi-major axes of the ellipses in Fig.~2(a).  We show in Fig.~3(a) the
separate effects of reducing the top quark mass errors to $\Delta m_t = \pm
2~\g/c^2$ (solid curves) and the errors on $\seff$ to $\pm 0.00023$ (dashed
curves) [which would be the error if we neglected the scale factor and the
error on $\alpha(M_Z)$ mentioned earlier]. Note the similar shapes of the
dashed ellipses in Fig.~3(a) and the solid ellipses in Fig.~2(b). Improvements
in measurements of $\seff$ would have roughly the same impact as improvements
in measurements of $M_W$.  Consistency between the two determinations would
provide further confirmation of our assumption that $S_W \simeq S_Z$. 

As an example of what global fits could provide in the not-too-distant future,
we show in Fig.~3(b) the $S-T$ ellipses that would result from the
aforementioned improvements in errors in all of $M_W,~m_t$, and $\seff$. Only
with a combination of these improvements can one hope to shrink the error
ellipses in such a way as to achieve meaningful limits on the Higgs boson mass.
\bigskip

\centerline{\bf VII.  SUMMARY}
\bigskip

Electroweak measurements continue to be made with ever greater precision,
affording potential challenges to theory. Nonetheless, the electroweak theory
has continued to be remarkably resilient. Indications for new physics have come
and gone over the years, but now it appears that measurements are beginning to
exclude either a very heavy Higgs boson or point to some other physics (such as
further quark or lepton doublets) that would allow such a boson to exist.  It
appears that the indirect route to such a boson provided by these measurements
is an arduous one.  Fortunately, instruments for direct searches (such as LEP
II, the future Large Hadron Collider at CERN, and possibly an upgraded Tevatron
at Fermilab) will be available in coming years.  Meanwhile, significant
contributions continue to emerge from non-accelerator experiments, and for this
we particle physicists are profoundly grateful. 
\bigskip

I am indebted to H. Frisch, B. G. Levi, J. Pilcher, and C. Wagner for useful
discussions. This work was supported in part by the United States Department of
Energy under Grant No. DE FG02 90ER40560. 
\bigskip

\def \ajp#1#2#3{Am. J. Phys. {\bf#1}, #2 (#3)}
\def \apny#1#2#3{Ann. Phys. (N.Y.) {\bf#1}, #2 (#3)}
\def \app#1#2#3{Acta Phys. Polonica {\bf#1}, #2 (#3)}
\def \arnps#1#2#3{Ann. Rev. Nucl. Part. Sci. {\bf#1}, #2 (#3)}
\def \cmts#1#2#3{Comments on Nucl. Part. Phys. {\bf#1}, #2 (#3)}
\def \cn{Collaboration}
\def \cp89{{\it CP Violation,} edited by C. Jarlskog (World Scientific,
Singapore, 1989)}
\def \dpfa{{\it The Albuquerque Meeting: DPF 94} (Division of Particles and
Fields Meeting, American Physical Society, Albuquerque, NM, Aug.~2--6, 1994),
ed. by S. Seidel (World Scientific, River Edge, NJ, 1995)}
\def \dpff{{\it The Fermilab Meeting: DPF 92} (Division of Particles and Fields
Meeting, American Physical Society, Batavia, IL., Nov.~11--14, 1992), ed. by
C. H. Albright \ite~(World Scientific, Singapore, 1993)}
\def \efi{Enrico Fermi Institute Report No. EFI}
\def \epl#1#2#3{Europhys.~Lett.~{\bf #1}, #2 (#3)}
\def \f79{{\it Proceedings of the 1979 International Symposium on Lepton and
Photon Interactions at High Energies,} Fermilab, August 23-29, 1979, ed. by
T. B. W. Kirk and H. D. I. Abarbanel (Fermi National Accelerator Laboratory,
Batavia, IL, 1979}
\def \hb87{{\it Proceeding of the 1987 International Symposium on Lepton and
Photon Interactions at High Energies,} Hamburg, 1987, ed. by W. Bartel
and R. R\"uckl (Nucl. Phys. B, Proc. Suppl., vol. 3) (North-Holland,
Amsterdam, 1988)}
\def \ib{{\it ibid.}~}
\def \ibj#1#2#3{~{\bf#1}, #2 (#3)}
\def \ichep72{{\it Proceedings of the XVI International Conference on High
Energy Physics}, Chicago and Batavia, Illinois, Sept. 6 -- 13, 1972,
edited by J. D. Jackson, A. Roberts, and R. Donaldson (Fermilab, Batavia,
IL, 1972)}
\def \ijmpa#1#2#3{Int. J. Mod. Phys. A {\bf#1}, #2 (#3)}
\def \ite{{\it et al.}}
\def \jpb#1#2#3{J.~Phys.~B~{\bf#1}, #2 (#3)}
\def \lkl87{{\it Selected Topics in Electroweak Interactions} (Proceedings of
the Second Lake Louise Institute on New Frontiers in Particle Physics, 15 --
21 February, 1987), edited by J. M. Cameron \ite~(World Scientific, Singapore,
1987)}
\def \ky85{{\it Proceedings of the International Symposium on Lepton and
Photon Interactions at High Energy,} Kyoto, Aug.~19-24, 1985, edited by M.
Konuma and K. Takahashi (Kyoto Univ., Kyoto, 1985)}
\def \mpla#1#2#3{Mod. Phys. Lett. A {\bf#1}, #2 (#3)}
\def \nc#1#2#3{Nuovo Cim. {\bf#1}, #2 (#3)}
\def \np#1#2#3{Nucl. Phys. {\bf#1}, #2 (#3)}
\def \pisma#1#2#3#4{Pis'ma Zh. Eksp. Teor. Fiz. {\bf#1}, #2 (#3) [JETP Lett.
{\bf#1}, #4 (#3)]}
\def \pl#1#2#3{Phys. Lett. {\bf#1}, #2 (#3)}
\def \pla#1#2#3{Phys. Lett. A {\bf#1}, #2 (#3)}
\def \plb#1#2#3{Phys. Lett. B {\bf#1}, #2 (#3)}
\def \pr#1#2#3{Phys. Rev. {\bf#1}, #2 (#3)}
\def \prc#1#2#3{Phys. Rev. C {\bf#1}, #2 (#3)}
\def \prd#1#2#3{Phys. Rev. D {\bf#1}, #2 (#3)}
\def \prl#1#2#3{Phys. Rev. Lett. {\bf#1}, #2 (#3)}
\def \prp#1#2#3{Phys. Rep. {\bf#1}, #2 (#3)}
\def \ptp#1#2#3{Prog. Theor. Phys. {\bf#1}, #2 (#3)}
\def \ptwaw{Plenary talk, XXVIII International Conference on High Energy
Physics, Warsaw, July 25--31, 1996}
\def \rmp#1#2#3{Rev. Mod. Phys. {\bf#1}, #2 (#3)}
\def \rp#1{~~~~~\ldots\ldots{\rm rp~}{#1}~~~~~}
\def \si90{25th International Conference on High Energy Physics, Singapore,
Aug. 2-8, 1990}
\def \slc87{{\it Proceedings of the Salt Lake City Meeting} (Division of
Particles and Fields, American Physical Society, Salt Lake City, Utah, 1987),
ed. by C. DeTar and J. S. Ball (World Scientific, Singapore, 1987)}
\def \slac89{{\it Proceedings of the XIVth International Symposium on
Lepton and Photon Interactions,} Stanford, California, 1989, edited by M.
Riordan (World Scientific, Singapore, 1990)}
\def \smass82{{\it Proceedings of the 1982 DPF Summer Study on Elementary
Particle Physics and Future Facilities}, Snowmass, Colorado, edited by R.
Donaldson, R. Gustafson, and F. Paige (World Scientific, Singapore, 1982)}
\def \smass90{{\it Research Directions for the Decade} (Proceedings of the
1990 Summer Study on High Energy Physics, June 25--July 13, Snowmass, Colorado),
edited by E. L. Berger (World Scientific, Singapore, 1992)}
\def \tasi90{{\it Testing the Standard Model} (Proceedings of the 1990
Theoretical Advanced Study Institute in Elementary Particle Physics, Boulder,
Colorado, 3--27 June, 1990), edited by M. Cveti\v{c} and P. Langacker
(World Scientific, Singapore, 1991)}
\def \waw{XXVIII International Conference on High Energy
Physics, Warsaw, July 25--31, 1996}
\def \yaf#1#2#3#4{Yad. Fiz. {\bf#1}, #2 (#3) [Sov. J. Nucl. Phys. {\bf #1},
#4 (#3)]}
\def \zhetf#1#2#3#4#5#6{Zh. Eksp. Teor. Fiz. {\bf #1}, #2 (#3) [Sov. Phys. -
JETP {\bf #4}, #5 (#6)]}
\def \zpc#1#2#3{Zeit. Phys. C {\bf#1}, #2 (#3)}
\def \zpd#1#2#3{Zeit. Phys. D {\bf#1}, #2 (#3)}

\end{document}